# Charge density wave in a band insulator

**Authors:** Md Shafayat Hossain[1,2]*†, Wenhao Liu[3]*, Yuqi Zhang[4,5,6]*, Qi Zhang[2], Chao Lei[7], Nana Shumiya[2], Kouta Dagnino[8], Maksim Litskevich[2], Yu-Xiao Jiang[2], Jia-Xin Yin[2], Nikhil Dhale[3], Zi-Jia Cheng[2], Byunghoon Kim[2], Yongkai Li[4,5,6], Tyler A. Cochran[2], Xian P. Yang[2], Fan Zhang[3], Yugui Yao[4,5], Zhiwei Wang[4,5,6]†, Bing Lv[3]†, Titus Neupert[8], Luis Balicas[9], M. Zahid Hasan[2]†

**Affiliations:**

[1]Department of materials Science and Engineering, University of California, Los Angeles, CA, USA.

[2]Laboratory for Topological Quantum Matter and Advanced Spectroscopy, Department of Physics, Princeton University, Princeton, New Jersey, USA.

[3]Department of Physics, University of Texas at Dallas, Richardson, Texas 75080, USA.

[4]Centre for Quantum Physics, Key Laboratory of Advanced Optoelectronic, Quantum Architecture and Measurement (MOE), School of Physics, Beijing Institute of Technology, Beijing 100081, China.

[5]Beijing Key Lab of Nanophotonics and Ultrafine Optoelectronic Systems, Beijing Institute of Technology, Beijing 100081, China.

[6]Material Science Center, Yangtze Delta Region Academy of Beijing Institute of Technology, Jiaxing 314011, China.

[7]Department of Physics, The University of Texas at Austin, Austin, TX 78712, USA.

[8]Department of Physics, University of Zurich, Winterthurerstrasse, Zurich, Switzerland.

[9]National High Magnetic Field Laboratory, Tallahassee, Florida 32310, USA.

†Corresponding authors, E-mail: shossain@seas.ucla.edu; zhiweiwang@bit.edu.cn; blv@utdallas.edu; mzhasan@princeton.edu.

*These authors contributed equally to this work.

## Abstract:

**Charge density wave (CDW) implies a periodic modulation of the charge density. Typically observed in metallic systems, CDWs arise from Fermi surface instabilities, resulting in the total or partial gapping of the Fermi surface. Here, we present experimental evidence for a CDW state emerging in a band *insulator* which has no Fermi surface. The bulk and surface of our material platform, α-$Bi_4Br_4$, is gapped over the entire Brillouin zone. Through topographic and spectroscopic imaging at low temperatures, we unveil an unexpected unidirectional charge modulation in α-$Bi_4Br_4$, breaking the lattice translation symmetry. The CDW develops at temperatures below 40 K and adds an energy gap atop the existing insulating gap of α-$Bi_4Br_4$. Furthermore, our transport measurements reveal nonlinear electrical conduction, a phenomenon conventionally associated with the sliding or phason mode of incommensurate CDWs. These highly unusual observations represent a new type of CDW and demand a new theoretical framework for CDWs.**

## Main text:

Fermi liquids are prone to charge density wave (CDW) transitions, especially in low-dimensional systems. In such systems, electrons and holes at the Fermi energy can experience a Fermi surface nesting, leading to a CDW transition[1,2]. This results in the opening of a spectral gap at the Fermi energy and at the nesting wavevector. Thus, electrons near the Fermi energy lower their overall energy. Frequently, the CDW transition occurs in parallel with a structural distortion, i.e., electron-phonon coupling, that stabilizes the nesting condition. However, for the transition to occur, the associated increase in elastic energy must be compensated by a more pronounced decrease in the electronic energy of the system. This was Peierls' explanation for the formation of a CDW in a one-dimensional atomic chain, leading to a metal-insulator transition[1,2]. Quasi-one-dimensional systems are particularly susceptible to this Fermi surface instabilities. In principle, this mechanism is not expected to apply to wavevectors connecting states away from the Fermi energy, given that the associated electronic interactions are expected to be weak to trigger a phase transition. Thus, CDW distortions are expected only at or very near the Fermi energy, and be associated with a Fermi surface instability. Additionally, beyond the Peierls' mechanism, a strong and wavevector-dependent electron-phonon coupling has been proposed to drive CDW transitions in quasi-2D materials[3-6]. In these CDWs, spectral reconstructions may occur across the entire electronic structure, suggesting that the CDW transition could represent a metal-metal Lifshitz transition at the Fermi surface[7]. Here, a CDW state opens a gap at its wavevector thus reconstructing the fermi surface and defining a new Brillouin zone. If the Fermi surface is only partially gapped, zone folding within the Brillouin zone defined by the new CDW periodicity leads to a deeply reconstructed remnant Fermi surface. Thus, according to the prevailing theories, a CDW requires the presence of a Fermi surface and typically instigates a metal-insulator or metal-metal transition at the Fermi surface. However, our discovery presents a departure from these conventional scenarios—an instance of a CDW spontaneously developing in absence of (charged) Fermi surface, leading to an insulator-insulator transition.

Our material platform is $\alpha$-$Bi_4Br_4$, a quasi-one-dimensional compound that bears resemblance to the CDW transition scenario originally proposed by Peierls. Importantly, however, $\alpha$-$Bi_4Br_4$ is *not* a metal, but a fully gapped (with a gap spanning the entire Brillouin zone) insulator[8-14]. $\alpha$-$Bi_4Br_4$ (see Methods Sec. I for details on crystal growth) is, in fact, a higher-order topological insulator whose bulk and surfaces are also fully gapped, with the conducting states lying along the sample hinges allowed by symmetry and topology[10-14]. In this work, we investigate $\alpha$-$Bi_4Br_4$ using scanning tunneling microscopy (STM); see Methods Sec. II for details on the STM measurements. Equipped with high spatial and energy resolution, STM is a powerful technique for probing charge order at the atomic level[15-17]. Here, we employ this technique to visualize the charge order in $\alpha$-$Bi_4Br_4$ spectroscopically. The crystal structure of $\alpha$-$Bi_4Br_4$ is depicted in Fig. 1**a**. The crystal features an interlayer AB-type stacking along the *c*-axis (Fig. 1**a** upper panel), with quasi-one-dimensional Bi-Br chains running along the *b*-axis within each layer (Fig. 1**a** lower panel)[8-14]. The lattice in the *ab* monolayer breaks the mirror symmetry along the *a*-axis, distinguishing the A-type layer from the B-type layer[8-14]. For STM measurements on $\alpha$-$Bi_4Br_4$, we use mechanical cleaving to expose the *ab* plane of the crystal. Figure 1**b** shows an atomically resolved STM topographic image of a clean, defect-free A-type surface, acquired at $T = 40$ K. The fast Fourier transform (FFT) of the topographic image, illustrated in the lower panel of Fig. 1**b**, reveals sharp Bragg peaks ($Q_{Bragg}$). Remarkably, as the sample is cooled down to 4 K, we observe a pronounced electronic density modulation, indicative of a charge order. This is clearly visualized in the topographic image shown in Fig. 1**c**, measured over the same area. Correspondingly, the respective FFT image (Fig. 1**c** lower panel) features additional charge order peaks marked with $Q_{CO}$. Comparing $Q_{CO}$ with $Q_{Bragg}$, we find the corresponding wavevector to be $\pm(-(0.05 \pm 0.015)\, a^* + (0.085 \pm 0.01) b^*)$, where $a^*$ and $b^*$ denote the Bragg wavevectors along the *a*- and *b*-axis, respectively. The period of the charge order is approximately $\lambda_{CO} \sim 3$ nm.

After identifying the charge order through topographic imaging, we conduct a series of energy-resolved spectroscopic measurements in the same area at $T = 40$ K and $T = 4$ K. Tunneling differential conductance ($\mathrm{d}I/\mathrm{d}V$) spectra at $T = 4$ K reveal a significant insulating gap (where $\mathrm{d}I/\mathrm{d}V \sim 0$) of approximately $\simeq 260$ meV, consistent with previous tunneling measurements. For a detailed explanation of how we determined the energy gap from the $\mathrm{d}I/\mathrm{d}V$ spectra, we refer the reader to the Supplementary Information Sec. I and Supplementary Fig. 1. In contrast, $\mathrm{d}I/\mathrm{d}V$ spectra at $T = 40$ K display a spectroscopic gap of around $\simeq 225$ meV, indicating a noticeable change in the energy gap between both temperatures that exceeds the associated thermal broadening energy. We attribute this change in the energy gap to the emergence of the charge order; a detailed discussion on the temperature dependence of the energy gap follows below.

Next, we proceed with the spectroscopic imaging of the charge order. Figure 2 summarizes the corresponding measurements. In Fig. 2**a**, we present the topographic image of the area. Figures 2**b** and 2**c** portray the corresponding $\mathrm{d}I/\mathrm{d}V$ maps and their respective FFT images acquired at -300 mV (occupied side of the energy gap) and 30 mV (unoccupied side of the energy gap). Notably, these $\mathrm{d}I/\mathrm{d}V$ maps also reveal a pronounced charge modulation, consistent with the topographic image. Importantly, by tracing the charge modulation patterns, we observe that the $\mathrm{d}I/\mathrm{d}V$ maxima (minima) at -300 mV align with $\mathrm{d}I/\mathrm{d}V$ minima (maxima) at 30 mV. This observation indicates a spectroscopic contrast reversal upon adjusting the tip-sample bias from the conduction band to the valence band (Figs. 2**b** and 2**c**). Such a contrast switch is typically associated with the electronic nature of the charge order[17-20]; see Supplementary Information Sec. III for further details on spectroscopic contrast reversal and its implications.

Having detected the spectroscopic signature of the charge order across a large area, we narrow our focus to a small region identified by the topographic image in Fig. 3**a**. Subsequently, we gather spatially-resolved $\mathrm{d}I/\mathrm{d}V$ spectra over this region and extract the corresponding energy gap value from each spectrum. Figure 3**b** illustrates the resulting spatial distribution of the energy gap, commonly referred to as a gap map[21]. We observe that the energy gap also exhibits atomic correlations, as indicated by the presence of Bragg peaks in the FFT of the gap map (Fig. 3**b** lower panel). Notably, the energy gap displays charge-order-induced correlations, akin to those observed in the topographic image, manifested through gap modulation in the gap map. Furthermore, the FFT of both the topography and the gap map reveals similar peaks corresponding to the charge order. It is worth noting that while the magnitude of the $\mathrm{d}I/\mathrm{d}V$ spectra is influenced by both the sample and the tip, the energy gap primarily correlates with the electronic properties of the sample. Therefore, the spatial gap modulation associated with the charge order–observed in several charge-density wave compounds–has been interpreted as evidence for the electronic nature of the charge order[20,22]. Likewise, our observation of gap modulation associated with the charge order indicates an electronic nature for the charge order. Notably, we have also observed a domain wall separating two domains of charge order (Fig. 4). These two domains are phase-shifted by approximately 162 degrees. The presence of domain walls in our system implies the existence of at least two energetically degenerate ordered states: one with $Q_{\mathrm{CO}} = \pm[-(0.05 \pm 0.015)\, a^* + (0.085 \pm 0.01)b^* + q_z]$ and the other with $Q_{\mathrm{CO}} = \pm[(0.05 \pm 0.015)a^* - (0.085 \pm 0.01)b^* + q_z]$, where $a^*$ and $b^*$ represent the Bragg wavevectors along the *a*- and *b*-axes, respectively and $q_z$ denotes the CDW wavevector along the *c*-axis (that our STM measurements cannot capture). Since neither configuration is energetically preferred, different regions of the sample naturally adopt one or the other, resulting in the formation of domain walls between these degenerate CDW states. This behavior further suggests that the CDW state has a finite coherence length, with the domain structure likely governed by the distribution of disorder that pins the respective domains.

After detecting the charge order on the A-type surface, we also identified B-type layers in our STM measurements. Figure 5**a** schematically depicts the B-type monolayers, distinguished from the A-type layers by the direction of mirror-symmetry breaking along the $a$-axis. In Figs. 5**b** and 5**c**, we present the atomically resolved topographic image of the B-type surface acquired at $T = 40$ K and 4 K, respectively. Akin to the A-type surface, the B-type surface also exhibits a pronounced charge modulation (Fig. 5**c**).

Next, we investigate the temperature dependence of the charge-order. For this purpose, we conduct energy-resolved spectroscopic measurements in the same area at various temperatures ranging from 4 K to 40 K. The resulting tunneling $\mathrm{d}I/\mathrm{d}V$ spectra are illustrated in Fig. 5**d**. As discussed in the context of Fig. 1**d**, the $\mathrm{d}I/\mathrm{d}V$ spectra at $T = 4$ K reveal a large insulating gap of $\simeq 260$ meV. Upon gradual heating, we observe a decrease in the spectroscopic gap. This trend is depicted in Fig. 4**e**, where we plot the spectroscopic gap as a function of temperature. We observe a significant change in the gap between $T = 20$ K to 30 K. It is plausible that the charge order emerges within this temperature range, leading to the development of an additional gap coexisting with the insulating gap of $\alpha$-$Bi_4Br_4$, rendering the CDW transition to be a surprising, insulator-insulator transition.

Surprises continue in our transport measurements where we substantiate the bulk nature of the charge order. The sample resistance as a function of temperature is displayed in Figs. 6**a,b** (see Methods Sec. III for details on device fabrication and transport measurements). The sample shows an insulating transport, consistent with the transport measurements by Li et al.[23]. The Arrhenius plot in Fig. 6**b** clearly reveals a steep increase in slope near $T = 30$ K. This increase is consistent with the development of a charge order enhancing the overall energy gap. It is, therefore, possible that the transition temperature for the CDW state in α-$Bi_4Br_4$ is close to $T = 30$ K. Note that, the suppression of the activated behavior at temperatures below 20 K has been reported before[14] and is possibly due to the gapless hinge states effectively short-circuiting the bulk gap by providing additional channels for carrier conduction.

To investigate the charge order, we explore the motion of this order under a DC electric field. The charge order is expected to be pinned by the disorder potential created by impurities but can be de-pinned[24] under a sufficiently large electric field causing the charge order to slide. Consequently, the current-voltage characteristics should exhibit non-linear behavior indicative of a sliding phason mode[24,25,26]. To explore this possibility, we applied a DC current between the source and drain probes and measured the sample's differential resistance ($\mathrm{d}V/\mathrm{d}I$) using a standard lock-in technique. We also measured the DC voltage across the voltage probes and extracted the electric field using $E = V/L$ ($L$ is the distance between the voltage probes). Fig. 6**c** displays a plot of $\mathrm{d}V/\mathrm{d}I$ as a function of $E$ at various temperatures. At small values of $E$, $\mathrm{d}V/\mathrm{d}I$ remains constant for all temperatures. However, for $T \leq 30$ K, $\mathrm{d}V/\mathrm{d}I$ sharply decreases above a threshold electric field ($E_{\mathrm{th}}$), indicating the onset of collective phason current[1,2]. Importantly, this sliding or phason mode is typically associated with CDWs originating from Fermi surface instabilities[1,2], which are absent in an insulating material like $\alpha$-$Bi_4Br_4$. Nevertheless, our observation of non-linear conduction in transport measurements aligns with our STM findings, providing further support for the presence of a CDW at low temperatures. Incommensurate CDWs display typical values of $E_{\mathrm{th}}$~10 mV/cm for the electric threshold field[1], which is one order magnitude smaller than the values extracted by us for $\alpha$-$Bi_4Br_4$ which exceed 100 mV/cm at low temperatures (see Supplementary Information Sec. V for additional analysis). The larger threshold electric field might not be surprising given that $E_{\mathrm{th}}$ was previously found to roughly scale with the square of the CDW gap[27].

The temperature dependence of $E_{th}$, defined as the onset of the sharp decrease in d$V$/d$I$ with respect to $E$, is illustrated in Fig. 6**d**. We find no signature of $E_{th}$ above $T \simeq 30$ K. Overall, our electrical transport experiments, demonstrating a well-defined $E_{th}$ that is consistent with the presence of the translation symmetry breaking charge order and its temperature onset that agrees with the scanning tunneling microscopy observations, serve as bulk evidence for the existence of the bulk charge density wave in α-$Bi_4Br_4$.

In discussing the transport properties of $\alpha$-$Bi_4Br_4$, it is noteworthy to mention the findings of Li et al.[23] They observed a pronounced increase in the temperature-dependent resistance around 23 K and interpreted it as a potential CDW transition at that temperature. This interpretation is consistent with our STM observations. Intriguingly, Li et al.[23] also observed a gradual shift of this putative transition temperature from 23 K under ambient pressure to 6 K at $P = 2.3$ GPa, with the transition eventually disappearing at $P \geq 3.0$ GPa, preceding a superconducting transition observed at low temperatures under $= 3.8$ GPa. This prior observation, in conjunction with our STM results, suggests that the charge order weakens with increasing pressure and vanishes just before the onset of superconductivity. Pressure tends to increase the dispersion of the electronic bands by increasing the overlap between atomic orbitals or induce structural transitions. Either scenario could lead to the suppression of the CDW state. The resulting charge fluctuations could trigger pressure-induced superconductivity. Future studies will be crucial to uncover the exact mechanism by which external pressure suppresses the CDW state in α-$Bi_4Br_4$.

Furthermore, our observations hold important implications for the interplay between topology and correlation. This topic has garnered intense interest in recent years, especially in the context of Moiré quantum matter[28-30], Kagome lattice compounds[16,31-33], and Weyl semimetals[17,26,34,35]. In these materials, electronic correlations often manifest as charge-ordered or superconducting states. For example, the interplay between band topology, charge order, and superconducting states, as seen in $AV_3Sb_5$ (A ≡ K, Rb, and Cs) Kagome compounds[16,31-33], and the intertwining of Weyl points with charge-ordered states in compounds like $Ta_2Se_8I$[17,26,35] and $TaTe_4$[34], have opened new research avenues. Similarly, our observation of charge order in the higher-order topological insulator $\alpha$-$Bi_4Br_4$ turns it into a material platform that allows studying the interplay between two correlated orders, i.e., charge order and — under pressure — superconductivity[23], with topology. Other topological materials that show such an interplay (e.g., $AV_3Sb_5$ kagome compounds[16,32,33,35-41] suffer from a complicated fermiology, while $\alpha$-$Bi_4Br_4$ shows bulk insulating behavior, or under pressure, very low carrier density[23]. In addition, we find that the charge order further enhances the topological gap, making the ensuing effects more robust; see Supplementary Information Sec. IV for further details. At the same time, the topology protected by time-reversal and inversion symmetry[10] remains preserved in the presence of the charge order.

The observation of a CDW in $\alpha$-$Bi_4Br_4$— a material that is fully gapped over its entire Brillouin zone and therefore lacks a Fermi surface—is rather unexpected. This is because CDWs states typically arise from a Peierls like instability that results in either total or partial gapping of the Fermi surface in metallic systems. An intriguing explanation could involve the development of orbital order, or the emergence of a broken symmetry state characterized by localized and occupied orbitals forming a regular pattern. However, orbital order typically arises from an orbital degenerate parent state, where a static Jahn-Teller distortion[42] or many-body effects break the degeneracy by opening an energy gap[43]. In the case of $\alpha$-$Bi_4Br_4$, there is no clear evidence of orbital degeneracy, making this explanation unlikely[8-11].

Lastly, we discuss a possible scenario for the formation of the CDW in α-$Bi_4Br_4$. In a conventional CDW scenario for quasi-one-dimensional systems, opening a gap at the Fermi wave vector $\pm 2k_F$ decreases the overall electronic energy of the metallic system while releasing entropy upon cooling. Frequently, this mechanism is accompanied by a lattice distortion that slightly modifies the electronic structure at the Fermi level to improve the so-called nesting condition between nesting electron and hole Fermi surfaces. However, for the transition to occur, the increase in elastic energy due to the distortion must remain inferior to the overall decrease in electronic energy produced by the transition. According to our STM results, the magnitude of the gap in α-$Bi_4Br_4$ is rather uniform within the *ab*-plane. However, it is possible and even likely that its magnitude is lower at specific *k*-space locations within the *ac*- and *bc*-planes. Therefore, by developing an electronic modulation at an incommensurate wavevector $Q_{CO} = \pm(-(0.05 \pm 0.015)\,a^* + (0.085 \pm 0.01)b^*) + q_z)$ connecting specific *k*-space portions of the valence and conduction bands, where the gap might display a minimum, α-$Bi_4Br_4$ would increase its local gap size within these planes, therefore, reducing its overall electronic energy; see Supplementary Fig. 9. To our knowledge, such a scenario has not yet been proposed or verified for any insulating system.

A CDW ground state has been theoretically proposed, though not yet experimentally verified, in β-$In_2Se_3$—a ferroelectric semiconductor characterized by a band gap. In that system, specific Se displacements minimize the potential energy of a flat optical phonon branch, leading to several predicted CDW phases[44]. Similarly, our calculated phonon dispersion for monolayer $Bi_4Br_4$ reveals imaginary phonon modes at incommensurate vectors (Supplementary Information Sec. VII and Supplementary Fig. 8), in reasonable agreement with the experimentally observed CDW wavevector. This indicates that the CDW in α-$Bi_4Br_4$ may also be phonon-driven. However, unlike β-$In_2Se_3$, α-$Bi_4Br_4$ crystallizes in a centrosymmetric $C2/m$ phase and shows no evidence for ferroelectricity. Therefore, despite the potentially shared phonon-driven mechanism, the physical context of the CDWs in these two systems differ significantly.

We also note that there are systems, such as $BaIrO_3$[45], where there is localized density of states at the Fermi level, associated with disorder and crystallographic defects, which vanish once the CDW gap opens. In stark contrast, according to our STM study the material platform α-$Bi_4Br_4$ is clean (Supplementary Information Sec. VI and Fig. 7) and fully gapped across the entire Brillouin zone, with an insulating gap exceeding 200 meV and no states within the gap, well before the onset of CDW. The charge order develops below 40 K, adding an energy gap to the existing insulating gap. Here, the gap size is at least an order of magnitude larger than the transition temperature towards the CDW state. Consequently, the observation of a CDW in α-$Bi_4Br_4$ represents the formation of a CDW in a true band insulator state. Further theoretical investigations will be necessary to elucidate the mechanism underlying the emergence of this charge-ordered state.

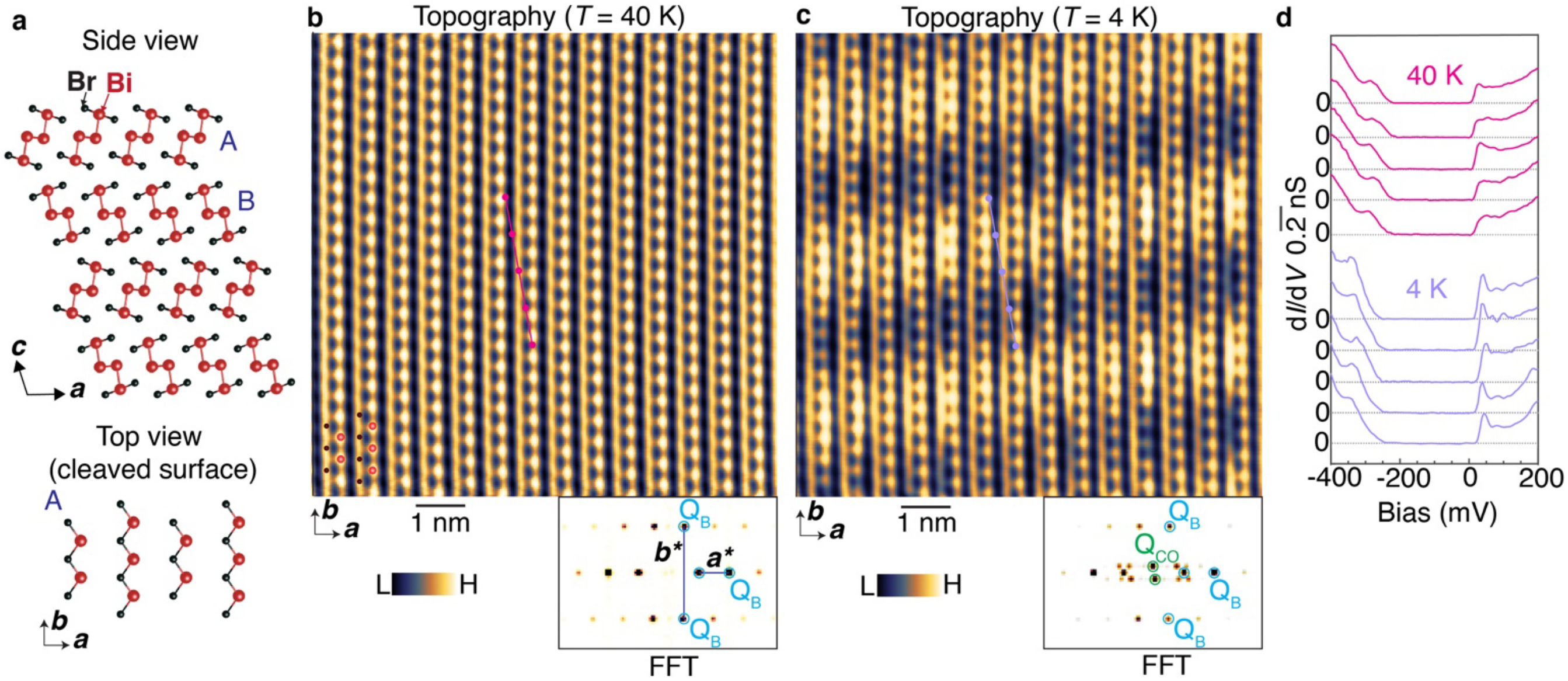

**Fig. 1: Observation of a charge order in $\alpha$-$Bi_4Br_4$. a,** Crystallographic structure of $\alpha$-$Bi_4Br_4$. Upper panel: side view of the $\alpha$-$Bi_4Br_4$ crystal showing the interlayer AB stacking. Lower panel: Top view of the A-type monolayers. **b, c**, Atomically resolved, STM topographic images measured at $T = 40$ K (panel **b**) and 4 K (panel **c**) on the same region of the A-type surface. H and L labels in the color bar denote high and low scale, respectively. The topography acquired at 4 K exhibits a pronounced charge modulation, which is absent at 40 K. The corresponding bottom panels display the Fourier transform images. The Bragg ($Q_{\mathrm{Bragg}}$) and the charge order ($Q_{\mathrm{CO}}$) peaks are indicated by blue and green circles, respectively. Tunneling junction setup for topographic images in panels **b** and **c**: $V$ = -300 mV, $I$ = 120 pA. **d,** Set of differential conductance (d$I$/d$V$) spectra acquired at $T = 40$ K and 4 K. Locations where the spectra were acquired are indicated by color-coded circles in the topographic images (panels **b** and **c**) measured at the corresponding temperatures. Tunneling junction setup: $V_{\mathrm{set}}$ = -600 mV, $I_{\mathrm{set}}$ = 420 pA, $V_{\mathrm{mod}}$ = 3 mV.

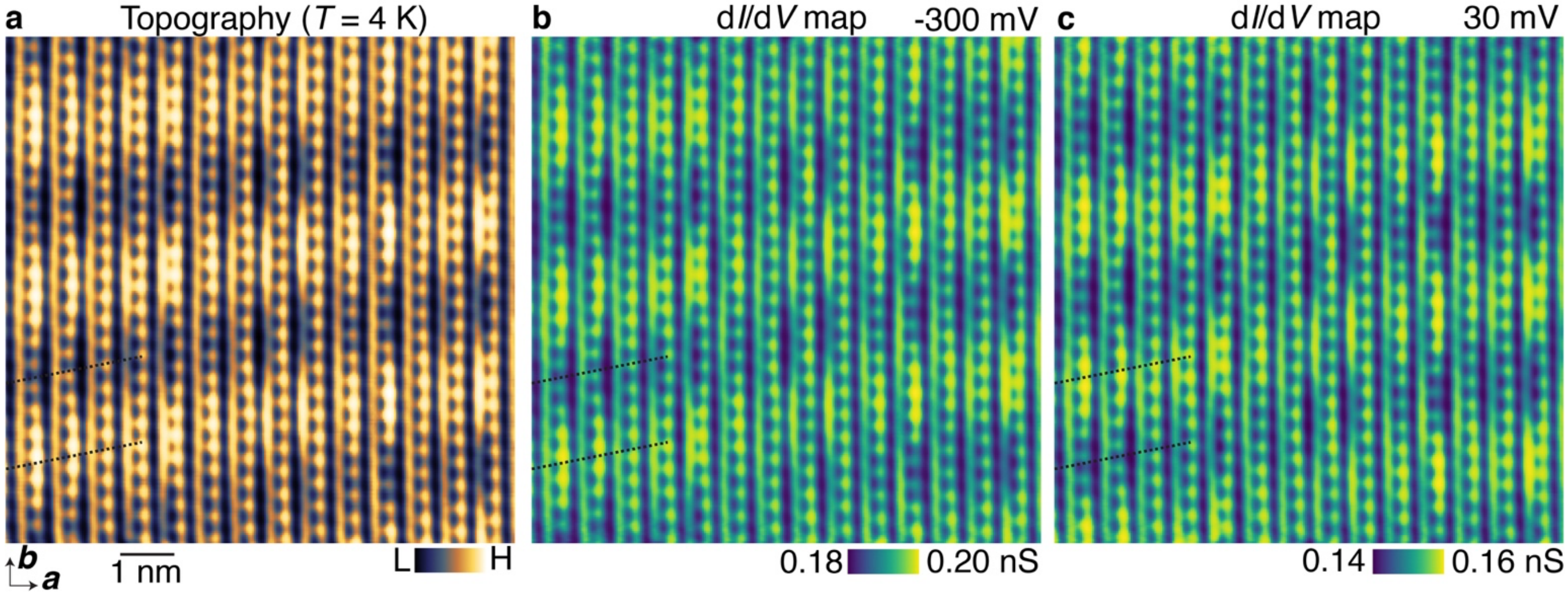

**Fig. 2: Spectroscopic signature of the charge order. a**, Same topographic image as in Fig. 1**c,** displaying a periodic charge modulation. Tunneling junction setup: $V$ = -300 mV, $I$ = 120 pA. **b**,**c**, Corresponding d$I$/d$V$ maps taken at -

300 mV (panel **b**) and 30 mV (panel **c**) revealing the same periodic charge modulation. The local density of states maxima (minima) at -300 mV correspond to the local density of states minima (maxima) at 30 mV. This is shown using two black dashed lines on the same location in the three panels and indicating the maxima and minima in the local density of states. The maxima and minima switch between -300 mV (panel **b**) and 30 mV (panel **c**). This observation signals a spectroscopic contrast reversal between the occupied and unoccupied sides of the gap, as expected for a charge ordered state. Tunneling junction setup for the d$I$/d$V$ maps: $V_{set}$ = -600 mV, $I_{set}$ = 420 pA, $V_{mod}$ = 5 mV.

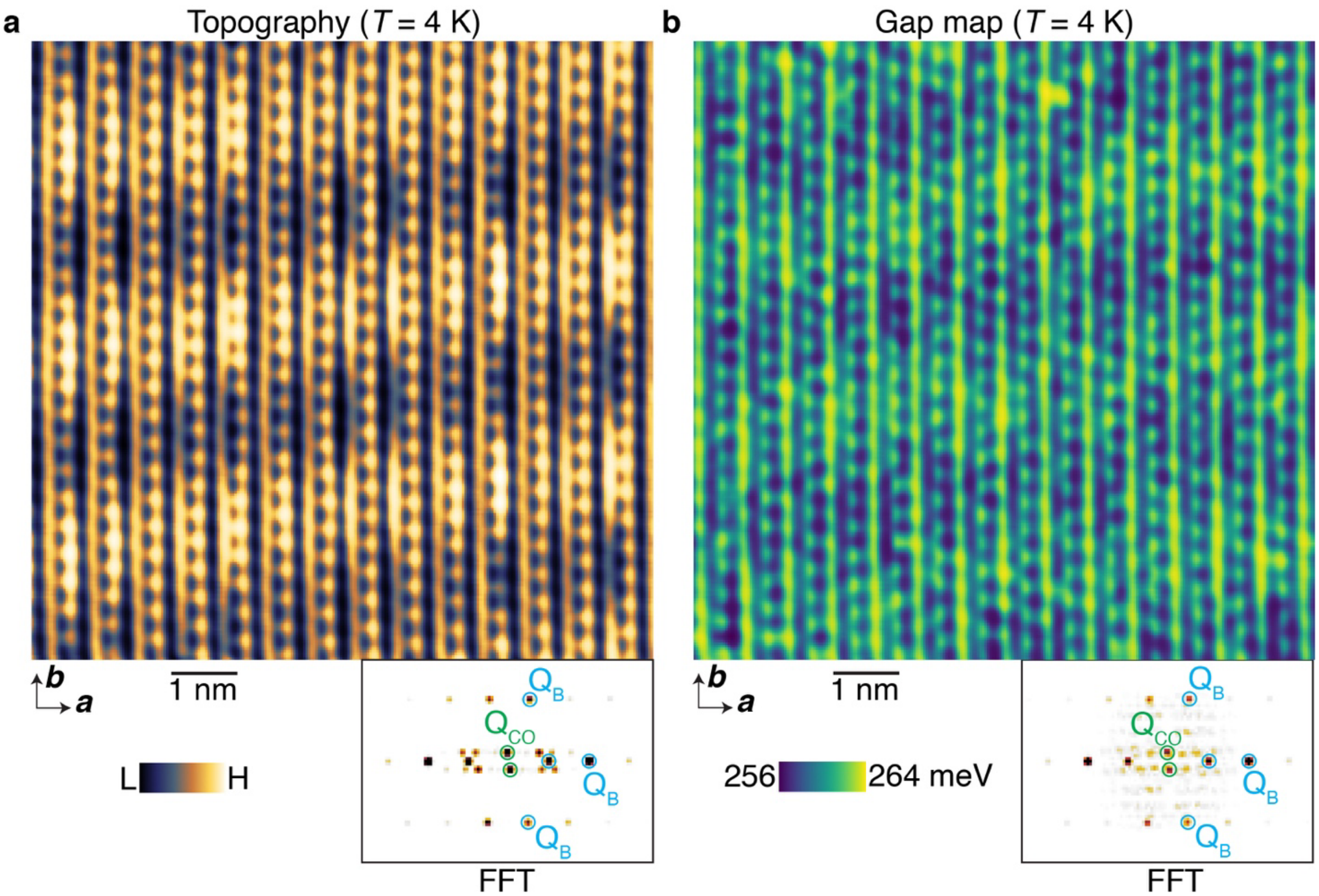

**Fig. 3: Manifestation of the charge order in the spectroscopic energy gap. a**, Same topographic image as in Fig. 1**c**, showing a pronounced charge modulation. Tunneling junction setup: $V$ = -300 mV, $I$ = 120 pA. **b**, Corresponding spectroscopic gap map, acquired at the same location, revealing a gap modulation associated with the charge order. Note that the charge density should be higher where the gap is smaller. Consequently, the protrusion in panel **a** is found in regions where the gap in panel **b** displays a depression and vice versa. Tunneling junction setup for the gap map: $V_{set}$ = -600 mV, $I_{set}$ = 420 pA, $V_{mod}$ = 2 mV. The bottom panels illustrate the corresponding Fourier transform images. Blue and green circles denote the locations of $Q_{CO}$ and $Q_{Bragg}$, respectively.

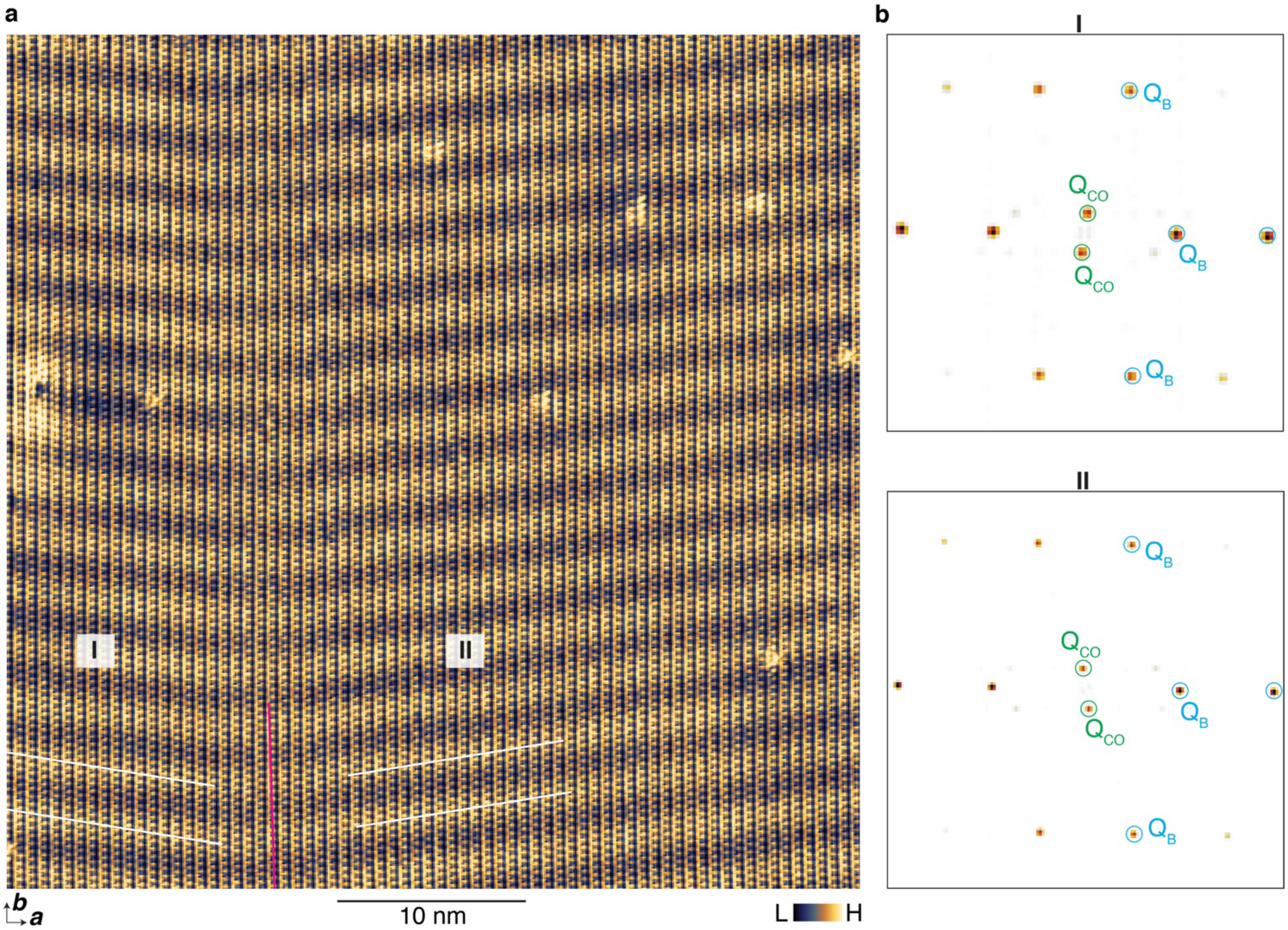

**Fig. 4: Observation of charge-ordered domains in $\alpha$-$Bi_4Br_4$. a,** STM topographic image acquired at $T = 4$ K displaying two charge-ordered domains, labeled as I and II, separated by a domain wall (indicated by a red line). The white lines denote the directions of the charge order in each domain. Tunneling junction setup: $V$ = -300 mV, $I$ = 120 pA. **b,** Fourier transform images of domains I (upper panel) and II (lower panel). Blue and green circles highlight the Bragg ($Q_{Bragg}$) and the charge order ($Q_{CO}$) peaks, respectively. The magnitude of the charge order wavevectors, $Q_{CO} = \pm[-(0.05 \pm 0.015)\,a^* + (0.085 \pm 0.01)b^*]$ and $Q_{CO}=\pm[(0.05 \pm 0.015)a^* - (0.085 \pm 0.01)b^*]$, appears to be the similar in the two domains. However, they are shifted by approximately 162 degrees in phase.

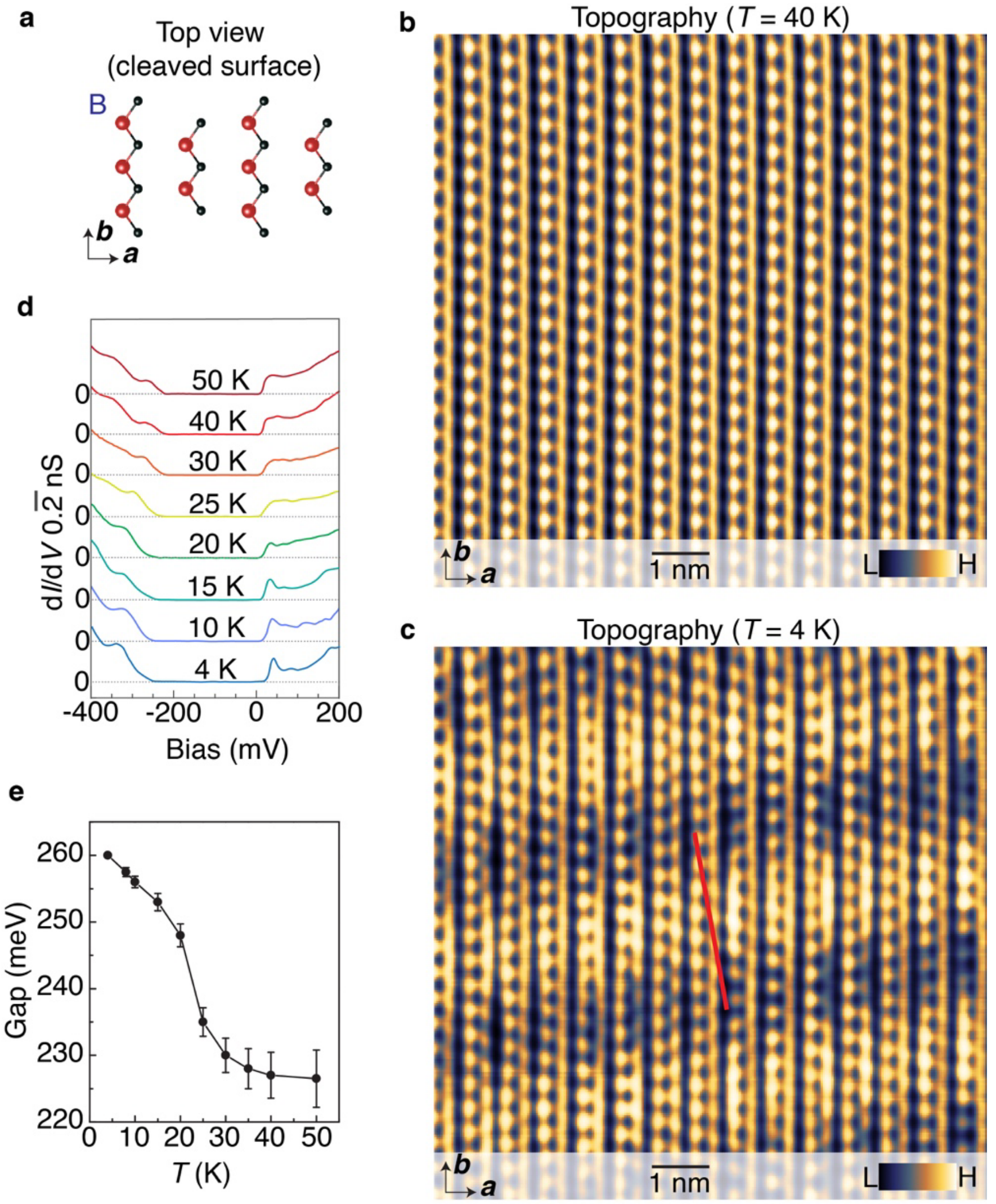

**Fig. 5: Temperature dependence of the charge order. a,** Crystallographic top view of the B-type $\alpha$-$Bi_4Br_4$ monolayers. **b, c**, Atomically resolved, STM topographic images measured at $T = 40$ K (panel **b**) and 4 K (panel **c**) on the same region in the B-type surface. The topography acquired at 4 K displays a pronounced charge modulation, which is absent at 40 K. Tunnelling junction set-up: $V = -300$ mV, $I = 120$ pA. **d,** Differential conductance (d$I$/d$V$) spectra acquired at different temperatures ranging from 4 K to 50 K. The spectrum for a given temperature is obtained via averaging the d$I$/d$V$ spectra measured along the red line shown in panel **c**. Tunneling junction setup: $V_{set} = -600$ mV, $I_{set} = 420$ pA, $V_{mod} = 3$ mV. **e,** Temperature dependence of the spectroscopic gap. The gap exhibits a clear enhancement at low temperatures which is likely associated with the emergence of the charge-order. The thermal broadening energy for each data point is indicated using vertical bars.

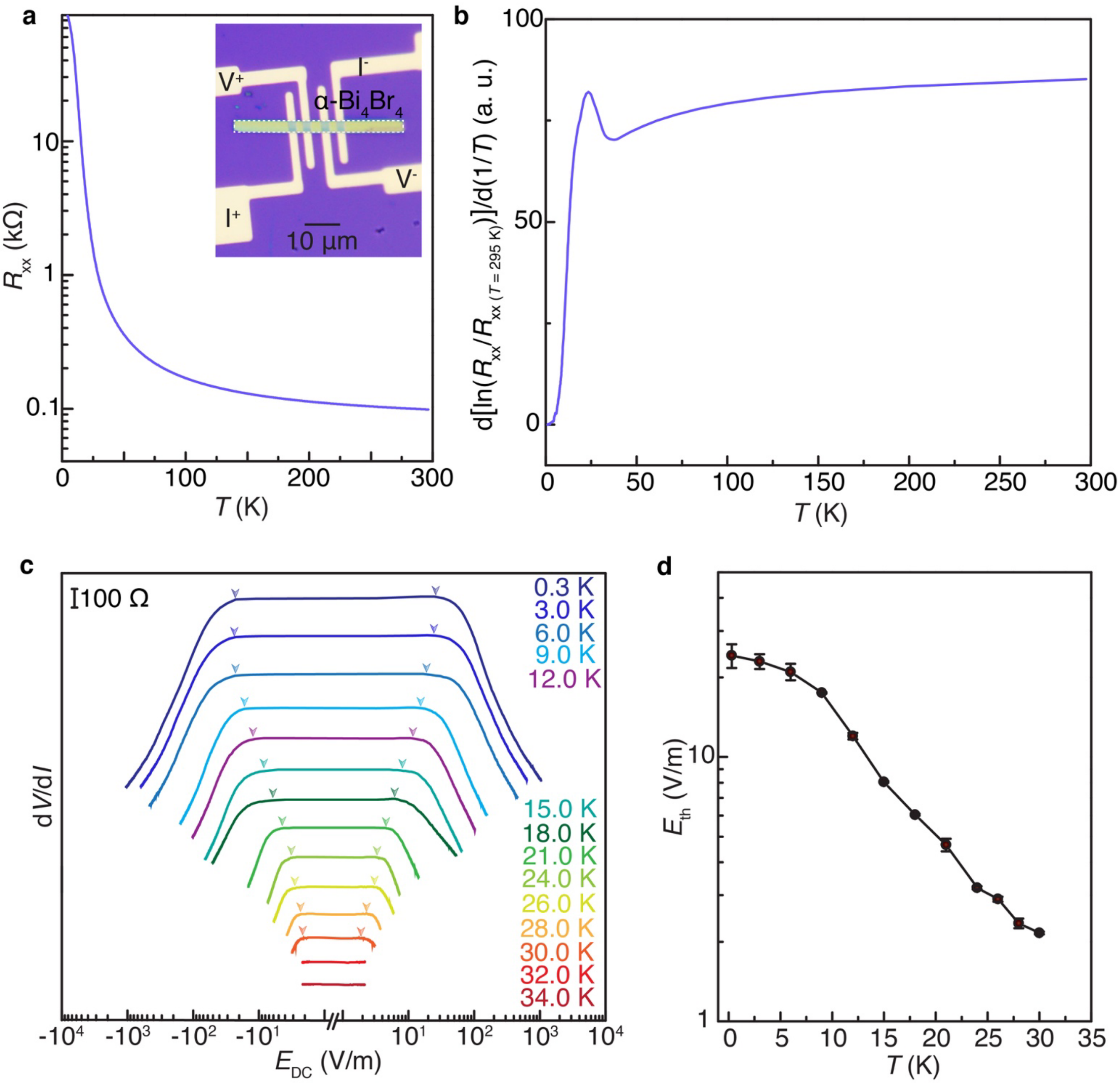

**Fig. 6: Transport evidence for the charge order in $\alpha$-$Bi_4Br_4$. a**, Four-probe resistance ($R_{xx}$) of the $\alpha$-$Bi_4Br_4$ device as a function of the temperature, $T$. The sample shows a clear insulating transport response. Inset: Optical microscopy image of the four-point probe device used for the transport measurements. The device comprises a mechanically exfoliated $\alpha$-$Bi_4Br_4$ flake. White dashed lines indicate the flake boundaries. Current ($I^+$ and $I^-$) and voltage ($V^+$ and $V^-$) probes for the four-terminal electrical transport measurements are also shown. **b**, d[ln($R_{xx}/R_{xx\,(T = 295\text{ K})}$)]/d(1/T) plotted as a function of T. The plot suggests an activated response, also revealing a steep increase in d[ln($R_{xx}/R_{xx\,(T = 295\text{ K})}$)]/d(1/$T$) near $T = 30$ K. **c**, Differential resistance d$V$/d$I$ as a function of the electric field at various temperatures. Traces for different temperatures are vertically offset for enhanced clarity. At low electric fields, d$V$/d$I$ remains constant for all temperatures. However, below the critical temperature $T_{CDW}$, d$V$/d$I$ sharply decreases above a threshold electric field $E_{th}$, indicating the onset of the collective phason mode or the electric field-

induced sliding of the translation symmetry breaking order. ($E_{th}$ is determined as the onset of the sharp decrease in d$V$/d$I$ with respect to the electric field. The arrows mark the locations of $E_{th}$ for each trace, both at positive and negative electric fields. **d**, Threshold electric field $E_{th}$ as a function of temperature. The error bars represent the variation between positive and negative $E_{th}$.

# Methods:

### I. Single crystal synthesis

The α-$Bi_4Br_4$ single crystals were synthesized via solid-state reactions. Initially, Bi (Alfa Aesar, 99.9999%) pieces were combined with $HgBr_2$ powders (Alfa Aesar, 99% +) in a stoichiometric Bi:Br ratio of 1:1. The mixture was then sealed in an evacuated quartz tube. Subsequently, the tube was placed into a horizontal two-zone tube furnace, with the raw materials positioned at the hot end at 265 °C and the cold end set at 210 °C. After several days, black needle-shaped crystals began to form at the cold end and in the middle of the quartz tube. To enhance crystal quality, the entire assembly underwent further annealing in a low-temperature oven at 160 °C for over one month following the initial two-week reaction period. This meticulous synthesis process yielded large crystals measuring up to $6 \times 0.7 \times 0.5$ $mm^3$, with needle-like morphology oriented along the $b$-axis and a substantial lateral area along the (001) surface. All procedures involving material handling and processing were conducted within a purified Ar-atmosphere glovebox, maintaining total $O_2$ and $H_2O$ levels below 0.1 ppm to ensure pristine conditions. The structural integrity of the single crystals was confirmed through X-ray single crystal diffraction, utilizing both a Bruker Apex DUO single crystal diffractometer and a Rigaku Smartlab X-ray diffractometer. Additionally, the composition was verified via SEM energy-dispersive X-ray spectroscopy (SEM-EDX) performed on a Zeiss EVO LS 15 SEM, utilizing an accelerating voltage of 20 keV.

### II. Scanning tunneling microscopy measurements

Scanning tunneling microscopy (STM) measurements were conducted on freshly cleaved samples. Single crystals underwent mechanical cleavage in situ at 77 K under ultra-high vacuum conditions ($< 5.0 \times 10^{-10}$ mbar) to achieve atomically resolved, pristine surfaces. Immediately after cleavage, the samples were inserted into the microscope head, pre-cooled to the $^4$He base temperature (4 K). For each cleaved crystal, surface areas over $5 \times 5$ $\mu m^2$ were explored to find atomically flat surfaces. Topographic images were acquired using a Unisoku Ir/Pt tip in constant current mode. Tunneling conductance spectra were obtained employing standard lock-in amplifier techniques with a lock-in frequency of 977 Hz. Details regarding the tunneling junction setup and modulation voltage for lock-in detection are provided in the corresponding figure captions. For temperature-dependent measurements, we retracted the tip from the sample and raised the temperature, stabilizing it for 12 hours before reapproaching the tip to the sample for tunneling measurements.

### III. Device fabrication for transport measurements

We used a polydimethylsiloxane stamp-based mechanical exfoliation technique to fabricate α-$Bi_4Br_4$ devices. We patterned the sample contacts on the silicon substrates with a 280 nm layer of thermal oxide using electron beam lithography, followed by chemical development and metal deposition (5 nm Cr and 35 nm Au). The fresh α-$Bi_4Br_4$ flakes were mechanically exfoliated from bulk single crystals on polydimethylsiloxane stamps. Before transferring them onto the $SiO_2$/Si substrates with pre-patterned Cr/Au electrodes, we identified suitable samples with good geometry using optical microscopy. Note that, we have used thick flakes (of the order of 50 nm) to capture the bulk properties, but in a uniformly thick sample. To preserve the intrinsic properties of the compound and minimize environmental effects, we encapsulated the samples using thin polymethyl methacrylate films with thicknesses around ~50 nm, which ensured that the samples on the devices were never exposed to air directly. All sample fabrication processes were performed in a glovebox with a gas purification system (<1 ppm of $O_2$ and $H_2O$).

**Data and materials availability:** All data needed to evaluate the conclusions in the paper are present in the paper. Additional data are available from the corresponding authors upon request.

**Author Contributions Statement:** STM experiments were performed by M.S.H. and N.S. Transport experiments were performed by M.S.H. and Q.Z. M.L. and Y.X.J. helped with the measurements. J.X.Y., Z.J.C., B.K., T.A.C., and X.P.Y. helped with the data interpretation. Crystal growth was carried out by W.L., Y.Z., Y.L., Z.W., N.D., Y.Y., and B.L. Theory calculations were performed by C.L., K.D., F.Z., and T.N. Figure development and writing of the paper were undertaken by M.S.H., L.B., T.N., C.L., and M.Z.H. M.Z.H. supervised the project. All the authors discussed the results, interpretation and conclusion.

**Competing interests:** The authors declare no competing interests.

**Acknowledgement:**
We acknowledge illuminating discussions with C. Yoon. M.Z.H. group acknowledges primary support from the US Department of Energy (DOE), Office of Science, National Quantum Information Science Research Centers, Quantum Science Center (at ORNL) and Princeton University; STM Instrumentation support from the Gordon and Betty Moore Foundation (GBMF9461) and the theory work; and support from the US DOE under the Basic Energy Sciences programme (grant number DOE/BES DE-FG-02-05ER46200) for the theory and sample characterization work including ARPES. L.B. is supported by the United States Department of Energy under the Basic Energy Sciences programme through grant no. DE-SC0002613. The National High Magnetic Field Laboratory acknowledges support from United States National Science Foundation cooperative agreement grant no. DMR-1644779 and the state of Florida. The crystal growth at the University of Texas at Dallas acknowledges the support by the US Air Force Office of Scientific Research (AFOSR) (FA9550-19-1-0037), National Science Foundation (NSF) (DMREF-2324033) and Office of Naval Research (ONR) (N00014-23-1-2020). T.N. acknowledges supports from the European Union's Horizon 2020 research and innovation programme (grant no. ERC-StG-Neupert-757867-PARATOP). Crystal growth at Beijing Institute of Technology is supported by the National Science Foundation of China (NSFC) (Grant No: 92065109), the National Key R&D Program of China (Grant Nos: 2020YFA0308800, 2022YFA1403400), the Beijing Natural Science Foundation (Grant No. Z210006), Y.G.Y. is supported by the National Science Foundation of China (NSFC) (Grant Nos: 12321004, 12234003). Z.W. thanks the Analysis & Testing Center at BIT for assistance in facility support. M.S.H. acknowledges support from Samueli Foundation.